\documentclass[conference]{IEEEtran}
\IEEEoverridecommandlockouts
\usepackage{cite}
\usepackage{amsmath,amssymb,amsfonts}
\usepackage{algorithmic}
\usepackage{graphicx}
\usepackage{textcomp}
\usepackage{multirow}
\usepackage{stfloats}
\usepackage{soul}
\usepackage[svgnames]{xcolor}
\usepackage{tikz}
\usetikzlibrary{trees}
\usepackage{svg}
\usepackage{listings}
\usepackage[most]{tcolorbox}
\usepackage{colortbl}
\usepackage{booktabs}
\usepackage{adjustbox}
\usepackage{needspace}
\usepackage[colorlinks=true, linkcolor=blue, urlcolor=magenta, citecolor=green]{hyperref}

\usepackage{balance}

\lstdefinelanguage{Verilog}{
  morekeywords={
    module, endmodule, input, output, wire, reg, parameter, localparam,
    always, begin, end, if, else, for, assign, case, endcase,
    default, @(posedge, @(negedge, initial, $fwrite, $fopen, $fclose,
    $display, $finish, $readmemh
  },
  sensitive=true,
  morecomment=[l]{//},
  morestring=[b]",
}

\lstdefinestyle{verilogstyle}{
  language=Verilog,
  basicstyle=\ttfamily\footnotesize,
  keywordstyle=\color{blue},
  commentstyle=\color{gray},
  stringstyle=\color{red!70!black},
  showstringspaces=false,
  breaklines=true,
  breakatwhitespace=false,
  frame=none,
}

\newif\ifzapst
\zapsttrue
\ifzapst
  \renewcommand{\st}[1]{} % Zap mode: delete content
\else
  \renewcommand{\st}[1]{\sout{#1}} % Normal mode: strikethrough
\fi
\def\BibTeX{{\rm B\kern-.05em{\sc i\kern-.025em b}\kern-.08em
    T\kern-.1667em\lower.7ex\hbox{E}\kern-.125emX}}
\begin{document}

\title{ArchXBench: A Complex Digital Systems Benchmark Suite for LLM Driven RTL Synthesis}
\iffalse
\title{Conference Paper Title*\\
{\footnotesize \textsuperscript{*}Note: Sub-titles are not captured in Xplore and
should not be used}
\thanks{Identify applicable funding agency here. If none, delete this.}
}
\fi 

\author{
    \IEEEauthorblockN{Suresh Purini\IEEEauthorrefmark{1}, Siddhant Garg\IEEEauthorrefmark{1}, Mudit Gaur\IEEEauthorrefmark{1}, Sankalp Bhat\IEEEauthorrefmark{1}, Sohan Mupparapu\IEEEauthorrefmark{1}, Arun Ravindran\IEEEauthorrefmark{2}}
    \thanks{\IEEEauthorrefmark{1}Computer Systems Group, International Institute of Information Technology, Hyderabad, India. Email: suresh.purini@iiit.ac.in, \{siddhant.garg, mudit.gaur, sankalp.b, sohan.mupparapu\}@research.iiit.ac.in}
     \thanks{\IEEEauthorrefmark{2}Department of Electrical and Computer Engineering, The University of North Carolina, Charlotte, USA. Email: arun.ravindran@charlotte.edu}
    % \thanks{\IEEEauthorrefmark{3}Computer Systems Group, IIIT, Hyderabad, India. Email: suresh.purini@iiit.ac.in}
}
%\author{\IEEEauthorblockN{Anonymous Authors}
%}
\iffalse
\author{\IEEEauthorblockN{1\textsuperscript{st} Given Name Surname}
\IEEEauthorblockA{\textit{dept. name of organization (of Aff.)} \\
\textit{name of organization (of Aff.)}\\
City, Country \\
email address or ORCID}
\and
\IEEEauthorblockN{2\textsuperscript{nd} Given Name Surname}
\IEEEauthorblockA{\textit{dept. name of organization (of Aff.)} \\
\textit{name of organization (of Aff.)}\\
City, Country \\
email address or ORCID}
\and
\IEEEauthorblockN{3\textsuperscript{rd} Given Name Surname}
\IEEEauthorblockA{\textit{dept. name of organization (of Aff.)} \\
\textit{name of organization (of Aff.)}\\
City, Country \\
email address or ORCID}
\and
\IEEEauthorblockN{4\textsuperscript{th} Given Name Surname}
\IEEEauthorblockA{\textit{dept. name of organization (of Aff.)} \\
\textit{name of organization (of Aff.)}\\
City, Country \\
email address or ORCID}
\and
\IEEEauthorblockN{5\textsuperscript{th} Given Name Surname}
\IEEEauthorblockA{\textit{dept. name of organization (of Aff.)} \\
\textit{name of organization (of Aff.)}\\
City, Country \\
email address or ORCID}
\and
\IEEEauthorblockN{6\textsuperscript{th} Given Name Surname}
\IEEEauthorblockA{\textit{dept. name of organization (of Aff.)} \\
\textit{name of organization (of Aff.)}\\
City, Country \\
email address or ORCID}
}
\fi 
\maketitle
%\IEEEpubid{979-8-3315-3762-3/25/\$31.00~\copyright~2025 IEEE}
%\IEEEpubidadjcol
\begin{abstract}
Modern SoC datapaths include deeply pipelined, domain-specific accelerators, but their RTL implementation and verification are still mostly done by hand. While large language models (LLMs) exhibit advanced code-generation abilities for programming languages like Python, their application to Verilog-like RTL remains in its nascent stage. This is reflected in the simple arithmetic and control circuits currently used to evaluate generative capabilities in existing benchmarks. In this paper, we introduce ArchXBench, a six-level benchmark suite that encompasses complex arithmetic circuits and other advanced digital subsystems drawn from domains such as cryptography, image processing, machine learning, and signal processing. Architecturally, some of these designs are purely combinational, others are multi-cycle or pipelined, and many require hierarchical composition of modules. For each benchmark, we provide a problem description, design specification, and testbench, enabling rapid research in the area of LLM-driven agentic approaches for complex digital systems design.

Using zero-shot prompting with Claude Sonnet 4, GPT 4.1, o4-mini-high, and DeepSeek R1 under a pass@5 criterion, we observed that o4-mini-high successfully solves the largest number of benchmarks, 16 out of 30, spanning Levels 1, 2, and 3. From Level 4 onward, however, all models consistently fail, highlighting a clear gap in the capabilities of current state-of-the-art LLMs and prompting/agentic approaches.

\end{abstract}


\section{Introduction}
\label{sec:introduction}

Designing a modern system-on-chip (SoC) datapath remains a largely manual and labor-intensive process. Engineers must translate high-level algorithmic intent into register-transfer-level (RTL) code, iteratively refine microarchitectures to meet area, frequency, and power targets, and verify functionality through simulation and formal checks. Each iteration in this flow, spanning specification analysis, microarchitecture planning, RTL implementation, synthesis, and verification, can expose new timing or resource violations, forcing redesigns and extending time-to-market.
As today’s SoCs integrate sophisticated cryptographic, signal-processing, image-processing, and machine-learning accelerators, designers must reason about deep pipelines, hierarchical reuse, and fine-grained parameter tuning, all of which greatly increase design complexity and effort.

\medskip
\noindent
\textbf{LLM opportunity.}  
The emergence of large language models (LLMs) has opened new opportunities for automating aspects of both software and hardware design. Recent LLMs have demonstrated the ability to generate source code directly from natural language specifications, motivating their application to HDL code generation \cite{aivril} \cite{autochip} \cite{mage} \cite{paradigm} \cite{rtlfixer} \cite{veriassist} \cite{verilogcoder} \cite{verigen}. This has led to a rapid expansion of research on LLM-based HDL synthesis, with efforts focusing on developing standardized benchmarks, specialized models, fine-tuning methods, and prompting strategies to improve code correctness and domain-specific reasoning. Additionally, techniques such as iterative feedback, multi-agent collaboration, and expert model ensembles have been explored to enhance the quality of generated code and tackle complex circuit descriptions.

\medskip
\noindent
\textbf{Limitations of existing benchmarks.}  
Current Verilog generation evaluation suites such as \textit{RTLLM}~\cite{rtllm}, and \textit{VerilogEval}~\cite{verilogeval-v2} contain tens of designs, but the majority are single-function arithmetic or control blocks.  
They seldom feature hierarchical composition, deep pipelining, or complex domain-specific accelerators, and therefore do not stress the architecture-level trade-offs that dictate quality of results (QoR) in real SoC flows.  
As a consequence, research on LLM-driven RTL generation risks over-fitting to toy problems and under-estimating the algorithmic, integration, and verification challenges that arise in practice.

\medskip
\noindent
\textbf{ArchXBench.}  
To close this realism gap, we introduce \emph{ArchXBench}, a six-level benchmark suite whose breadth and depth mirror the datapath diversity found in production SoCs (see Table~\ref{tab:archxbench_summary}).  
Beginning with ripple-carry adders and progressing to full AES cores, CNN convolutions, and streaming FFT pipelines, ArchXBench captures:  
(i)~combinational, multi-cycle, and deeply pipelined implementations of the \emph{same} functionality;  
(ii)~hierarchical and parametric constructions that expose latency–area trade-offs; and  
(iii)~application domains—cryptography, signal and image processing, and machine learning—that dominate contemporary accelerator roadmaps.

\medskip
\noindent
\textbf{Alignment with the SoC design flow.} 
Each benchmark directory contains a problem description in natural language, an RTL-like interface specification, and a Verilog testbench; Levels~5--6 additionally provide Python reference models that generate stimuli and golden outputs.  
This structure enables researchers to investigate every AI-assisted design step:  
\emph{specification comprehension} from unstructured text;  
\emph{micro-architecture planning} through parameter exploration (e.g., unroll factors, block sizes);  
\emph{RTL generation} across increasingly large codebases;  
and \emph{verification} against cycle-accurate models.  
Because designs scale from a few hundred to tens of thousands of lines of RTL, ArchXBench highlights how LLM prompting, fine-tuning, retrieval, or search strategies must evolve as complexity increases.

\medskip
\iffalse
\noindent
\textbf{Extensibility.}  
ArchXBench is deliberately open-ended.  
New vertical layers can attach implementation scripts—for logic synthesis, power estimation, or physical design—enabling future tasks such as \emph{spec $\rightarrow$ timing-clean gate-level netlist}.  
Horizontally, additional domains (e.g.\ networking or storage) can be inserted without disrupting existing automation harnesses.  
A clear directory schema, plus automated scoring pipelines, encourages community contributions and leaderboards that track both functional correctness and QoR metrics.
\fi 
%\medskip
\noindent
\textbf{Impact.}  
ArchXBench provides the research community with a well-defined testbed of realistic yet self-contained datapath challenges. These benchmarks support the development and evaluation of advanced AI techniques—such as model finetuning, prompting, in-context learning, retrieval-augmented generation, agent-based methods, program synthesis, and formal verification—targeted at solving complex, real-world RTL design problems.

ArchXBench is accessible at this public \href{https://github.com/sureshpurini/ArchXBench}{GitHub repository}. As our goal in this paper is to establish a simple baseline rather than conduct a comprehensive study on prompting or agentic approaches, we used straightforward zero-shot prompting under a pass@5 criterion. In our evaluation, we found that among Claude Sonnet 4, GPT 4.1, o4-mini-high, and DeepSeek R1, the o4-mini-high model successfully solved the largest number of benchmarks—16 out of 30—spanning Levels 1, 2, and 3. While all models consistently struggled from Level 4 onward, this highlights exciting opportunities for advancing LLM capabilities and developing improved prompting and agentic strategies to tackle increasingly complex hardware designs.

The remainder of the paper is structured as follows. Section \ref{sec:related_work} reviews existing Verilog benchmarks relevant to LLM-based hardware generation. Section \ref{sec:archbench} describes the design and components of ArchXBench. Section \ref{sec:evaluation} presents our initial evaluation of ArchXBench using state-of-the-art commercial LLMs. 
Section~\ref{sec:discussion} provides a high-level discussion on how LLMs handle the benchmark complexity and hints at the way forward for agentic approaches.  Finally, Section \ref{sec:conclusions} summarizes our contributions and outlines directions for future work.

\section{Related Work}
\label{sec:related_work}
This section presents a concise review of prior work relevant to benchmarking methodologies for LLM based HDL generation.

\subsection{Benchmarks for Verilog generation with LLMs}
To measure progress in HDL code generation, researchers have developed benchmark suites and evaluation frameworks tailored to this domain. One of the earliest efforts was by Thakur et al. \cite{thakur2023benchmarking}  who defined a set of Verilog programming tasks and evaluated various LLMs (GPT-3, GPT-J, etc.) on their functional correctness. 

The original \textit{VerilogEval-v1} dataset \cite{verilogeval-v1} includes two types of problem descriptions: VerilogEval-machine and VerilogEval-human. The VerilogEval-machine descriptions are automatically generated by GPT-3.5 based on the corresponding RTL solutions, whereas the VerilogEval-human descriptions are manually authored and sourced from the \textit{HDLBits} repository. \textit{VerilogEval-v2} \cite{verilogeval-v2} extends the original VerilogEval benchmark by supporting specification-to-RTL generation tasks in addition to code completion. It introduces in-context learning (ICL) prompts, fine-grained failure classification, and a Makefile-based infrastructure for scalable and flexible evaluation. These enhancements enable deeper analysis of LLM behavior, highlight variability across models and tasks, and provide better tools for prompt tuning and debugging Verilog code generation.

Lu et al. released \textit{RTLLM} \cite{rtllm}, an open-source benchmark for RTL generation with LLMs.  It contains 50 designs covering a wide range of complexities and scales covering Arithmetic, Control, Memory, and Miscellaneous circuits including signal generators, frequency dividers, and RISC-V core components. Each design includes a natural language description specifying the intended functionality, module name, and input/output signals; a testbench with multiple test cases for verifying functional correctness; and a human-written reference Verilog design. The benchmark defines three evaluation goals—syntax, functionality, and design quality—to systematically assess the correctness and quality of auto-generated RTL designs.

Batten et al. introduced \textit{PyHDL-Eval} \cite{pyhdl-eval}, a benchmark for evaluating LLMs on Python-based HDLs such as PyMTL3 and Amaranth. It includes parameterized design tasks, simulation-based correctness checks, and code quality metrics. Unlike Verilog- or VHDL-focused benchmarks, PyHDL-Eval targets high-level hardware modeling, enabling evaluation of LLMs in emerging Python-centric design flows.

The introduction of benchmarks such as VerilogEval, RTLLM, and others has been instrumental in measuring the progress of LLM-based RTL generation \cite{rtlfixer, veriassist, aivril, verilogcoder, mage, paradigm}. These benchmarks highlight the capabilities of current models—such as effective handling of small combinational logic—and expose their limitations, particularly with complex state machines or long-form code. By providing standardized evaluation frameworks, they allow researchers to monitor improvements and pinpoint areas where model architectures or prompt strategies require refinement. However, as noted in the introduction, these benchmarks provide limited coverage of hierarchical composition, deep pipelining, and domain-specific accelerators. As a result, they fall short of evaluating the architectural trade-offs that impact quality of results (QoR) in realistic system-on-chip (SoC) design flows.

\begin{table*}[t]
\centering
\caption{Summary of ArchXBench Benchmark Levels}
\renewcommand{\arraystretch}{1.2}  % default is 1.0
\begin{tabular}{|c|p{4.2cm}|p{9cm}|}
\hline
\textbf{Level} & \textbf{Design Focus} & \textbf{Representative Domains / Examples} \\
\hline
0 & Very simple digital circuits & Muxes, demuxes, encoders, decoders, counters, etc; included for completeness. \\
\hline
1a & Combinational and simple multi-cycle circuits & Ripple-carry adders, shift-and-add multipliers, LFSRs, barrel shifter, LUT-based AES S-box, etc. \\
\hline
1b & Parametric and hierarchical circuits & 32-bit RCA with 4-bit CLAs, shift-and-add multiplier with parametric unroll factor, etc. \\
\hline
1c & Advanced arithmetic units & Kogge-Stone and Brent-Kung adders, Booth and Wallace/Dadda multipliers, integer dividers, etc. \\
\hline
2 & Pipelined arithmetic and cryptographic components & Pipelined RCA and CLA adders, pipelined Wallace and Dadda multipliers, AES single round, etc.\\
\hline
3 & Floating-point Airthmetic and iterative algorithms & Floating-point adders and multipliers, gradient descent, Newton-Raphson method for computing the root of a polynomial, etc.\\
\hline
4 & Pipelined signal processing circuits & FFT, IFFT, low-pass, high-pass, and band-pass filters, etc. \\
\hline
5 & Image processing and ML blocks & 2D convolution, unsharp mask, Harris corner detection, systolic GEMM, etc. \\
\hline
6 & Full subsystems and integrated designs & AES encryption/decryption cores, 3D CNN convolution, streaming FFT/DCT, FIR/IIR filter pipelines, quantized matmul, multi-channel Conv2D, etc. \\
\hline
\end{tabular}
\label{tab:archxbench_summary}
\end{table*}

\section{ArchXBench Benchmark Description}
\label{sec:archbench}

ArchXBench is a hierarchical benchmark suite designed to support the development and evaluation of AI methods for hardware design, particularly approaches that leverage Large Language Models (LLMs) for synthesizing and optimizing data-path intensive digital systems. It explicitly focuses on computation-heavy subsystems and excludes control-dominated designs such as DMA engines, memory controllers, and interface circuitry. The benchmark emphasizes architectural diversity and computational complexity to foster research into automated RTL generation and performance optimization.

\subsection{Benchmark Levels and Domains}

The suite is organized into six distinct levels, with Level~1 further divided into three sublevels (1a, 1b, and 1c), yielding a total of 51 benchmark designs. An auxiliary Level~0 contains a set of very simple digital circuits, such as multiplexers, counters, encoders, and shift registers, included primarily for completeness. The benchmarks span a wide range of architectural complexities, from simple combinational circuits to multi-cycle iterative and pipelined systems, and include hierarchical compositions of such units. The organization captures critical trade-offs among area, latency, throughput, and power, thereby facilitating comprehensive architectural design space exploration.

The classification is primarily based on architectural features, but an alternative application-oriented perspective is also supported. Higher-level benchmarks in particular align with key domains such as cryptography, signal processing, image processing, and machine learning. This dual classification enables both fundamental architecture-driven analysis and domain-specific research in emerging areas.

Level~1a contains elementary combinational circuits such as ripple carry and carry look-ahead adders, along with simple multi-cycle designs like shift-and-add multipliers and linear feedback shift registers. It also includes basic components of cryptographic systems, including AES S-boxes implemented as LUTs and $\mathrm{GF}(2^8)$ multipliers. Level~1b introduces hierarchical and parametric designs that permit architectural tunability. Representative designs include 32-bit ripple carry adders built from 4-bit carry-lookahead adder blocks, 32-bit carry-skip adders composed of 4-bit skip sections, and shift-and-add multipliers with configurable unroll factors. These benchmarks allow exploration of design trade-offs, as varying parameters such as block size or unroll factor can reduce latency at the expense of increased area and critical path length. Level~1c advances to complex arithmetic units such as Kogge-Stone (KS) and Brent-Kung (BK) adders, and Booth and Wallace multipliers. KS and BK adders are parallel-prefix designs known for their logarithmic carry propagation, balancing speed, area, and wiring complexity. Booth multipliers optimize signed multiplication through bit pattern encoding, while Wallace tree multipliers accelerate partial product summation using parallel reduction trees. The level also includes integer dividers, covering restoring architectures that iteratively compute quotient bits via comparison and subtraction.

Level~2 benchmarks extend earlier arithmetic circuits with pipelined architectures and introduce cryptographic components such as a single AES-128 encryption round. The inclusion of pipelined Wallace Tree and Dadda multipliers introduces new verification and timing challenges, such as pipeline stage balancing and clock domain closure. While pipelined ripple-carry and carry-lookahead adders are of reasonable complexity, the pipelined versions of Wallace Tree and Dadda multipliers are substantially more complex, due to their deep parallel reduction structures and dense interconnections, making their design and verification considerably more challenging. These designs mark a shift from latency-driven optimization to high-throughput hardware realization.

Level~3 benchmarks explore two primary themes: the implementation of IEEE-compliant floating-point arithmetic circuits, such as adders and multipliers, and the application of iterative fixed-point algorithms, such as the Newton-Raphson method for square root computation, the Newton-Raphson method for polynomial root finding, the Gauss-Seidel method for solving simultaneous linear equations, and gradient descent for optimizing quadratic polynomials. Here, the term \emph{fixed-point} carries a deliberate dual meaning --- it refers both to the finite-precision numeric representation used for computations and, in a mathematical sense, to the convergence of iterative methods toward a stable fixed point (i.e., a solution that remains unchanged under further iterations).

\begin{figure*}[t]
    \centering
    %\includesvg[scale=0.3]{figures/archxbench_corrected_diagram.svg}
    \includegraphics[
    trim=0 6cm 0 0, % trim margins (order is left, bottom, right, top)
  clip, % required for trimming to work
    width=\textwidth]{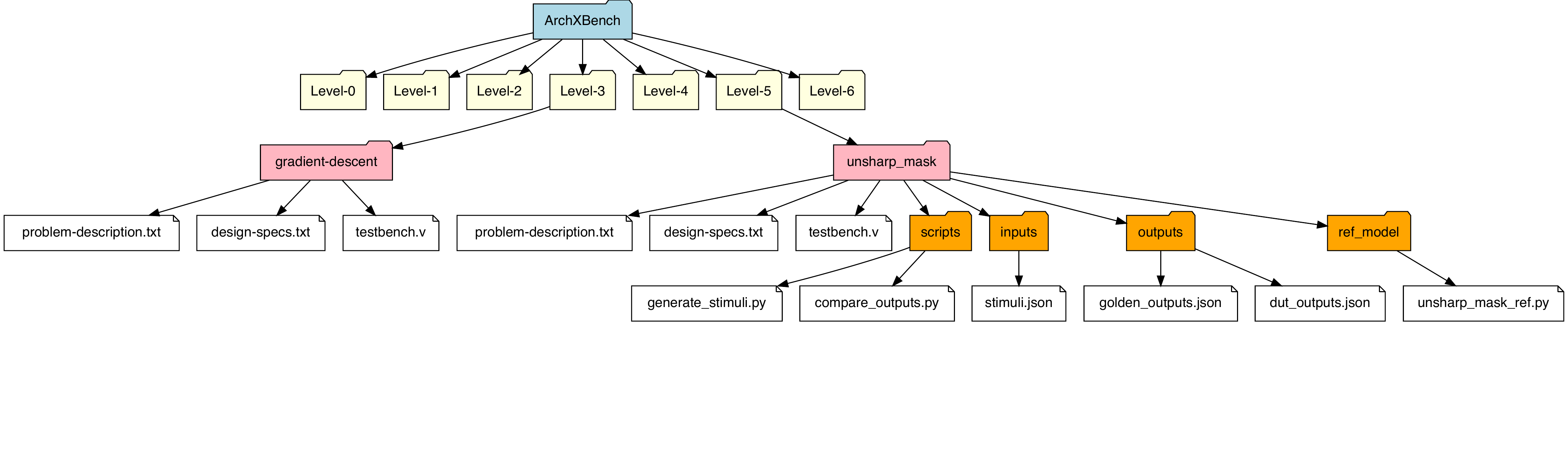}
    \caption{ArchXBench directory structure.}
    \label{fig:archxbench-diagram}
\end{figure*}

Level~4 advances to pipelined floating-point adders and multipliers and introduces iterative, multi-cycle implementations of signal processing benchmarks, such as FFT, IFFT, and low-pass, high-pass, and band-pass filters. At this stage, the focus is on architectural simplicity and resource efficiency, where operations are spread across multiple cycles to reduce hardware complexity. In contrast, more advanced pipelined versions of these signal processing benchmarks -- aimed at maximizing throughput and exploiting deep parallelism -- are introduced later in Level~6. 

Level~5 benchmarks represent full image processing pipelines and machine learning primitives. Examples include Unsharp Mask and Harris Corner Detection, which are expressed as directed acyclic graphs (DAGs) of stencil and pointwise operations. These designs rely on line buffers and hardware FIFOs to manage data reuse and inter-stage communication. This level also includes systolic array architectures for matrix multiplication, essential for ML inference engines such as TPUs. % as well as dynamic programming algorithms like Needleman-Wunsch used in bioinformatics.

Level~6 contains benchmarks with complex datapaths composed of multiple interconnected modules, reaching an architectural complexity comparable to domain-specific accelerators. These benchmarks include complete systems such as AES cores (with key expansion, encryption, and decryption), 3D convolutional blocks, multi-channel 2D convolution units for processing spatial and temporal data (e.g., in image and video applications), quantized matrix-multiplication units tailored for efficient neural network inference, and advanced FFT and FIR/IIR pipelines for signal processing. The convolutional benchmarks focus on accelerating core layers used in deep learning models, while the quantized matrix-multiplication benchmarks emphasize reduced-precision arithmetic to enhance computational efficiency and energy savings during inference. Together, these benchmarks challenge agentic synthesis tools with their complex, multi-module designs and advanced pipelining strategies.

\subsection{Directory Organization and Artifacts}
The \textbf{ArchXBench} benchmark suite is organized as a hierarchical directory structure, reflecting its multi-level classification of benchmarks, refer Figure~\ref{fig:archxbench-diagram}. At the top level, directories labeled \texttt{Level-0} through \texttt{Level-6} house progressively complex designs, ranging from simple arithmetic circuits to highly integrated systems. Each benchmark resides in its own subdirectory containing three key artifacts: a human-readable \texttt{problem-description.txt} outlining the design goals, background, constraints, and deliverables; a \texttt{design-specs.txt} file detailing the formal interface specification including inputs, outputs, and parameters; and a Verilog \texttt{testbench.v} for functional verification. Simpler benchmarks, such as \texttt{gradient-descent} under \texttt{Level-3}, follow this minimal three-file structure. In contrast, more advanced benchmarks, like \texttt{unsharp\_mask} under \texttt{Level-5}, include additional subdirectories such as \texttt{scripts} (for generating stimuli and comparing outputs), \texttt{inputs} (holding test input vectors), \texttt{outputs} (containing both golden reference and DUT results), and \texttt{ref\_model} (providing a Python-based functional reference implementation). This directory organization supports modularity, scalability, and reproducibility, ensuring that both straightforward and highly complex benchmarks can be systematically evaluated within the suite.

As an example, Listings 1-4 in Appendix A present the details of a Level 6 benchmark involving a 3D convolution accelerator. The complete ArchXBench benchmark suite is available at this public  \href{https://github.com/sureshpurini/ArchXBench}{GitHub repository}.

\st{
Each benchmark is housed in a dedicated directory containing three key artifacts. The \texttt{problem-description.txt} file provides a human-readable design specification, including the title, background, objectives, constraints, performance expectations, and deliverables. The \texttt{design-specs.txt} file describes the design interface in terms of inputs, outputs, and parameters. A Verilog testbench is provided for functional verification of the generated RTL.

For Levels~5 and~6, Python-based reference models are included to support high-level functional verification. These models contain scripts for generating input stimuli and for comparing the output of the synthesized RTL design against high-level specifications. This infrastructure is essential for complex systems where test vector generation and output verification must be tailored to the specific computational context of the benchmark.

In summary, ArchXBench offers a structured and extensible testbed for exploring the synthesis of complex digital systems. It spans a broad spectrum of architectural and domain-specific designs, enabling robust evaluation of LLM-driven and agentic techniques for RTL generation and optimization.
}

\section{Evaluation}
\label{sec:evaluation}
In this section, we evaluate the performance of state-of-the-art large language models (LLMs) on the ArchXBench benchmark suite. We consider four leading models: Sonnet 4.0 (Claude), GPT-4.1 and o4-mini-high (OpenAI), and DeepSeek R1 (DeepSeek). Using a straightforward prompting strategy, we provide each model with the corresponding \texttt{problem-description.txt} and \texttt{design-specs.txt} files and assess two key aspects: syntactic correctness and functional correctness. Additionally, for cases where the generated code is functionally correct, we examine whether it adheres to the architectural requirements specified in the benchmark. For example, if a task requires generating a pipelined multiplier circuit but the model produces an iterative design, we classify it as architecturally incorrect.

To account for the inherent variability in LLM outputs, we adopt a {\it pass@5} evaluation approach: we generate five independent completions (samples) per benchmark and consider the task successful if at least one of the five passes all evaluation criteria. This metric helps capture the model’s best-case performance under non-agentic prompting. Importantly, we do not apply any in-context examples, agentic techniques or multi-step refinement strategies in this evaluation; the goal here is to establish a baseline benchmark performance that can serve as a foundation for advancing research in LLM-driven agentic approaches for the architectural design of digital systems with highly complex datapaths.

Table~\ref{tab:sixtybenchmarks} summarizes the results for benchmarks from Levels~1 to~6. Up to Level~3, we provide Verilog reference designs, but beyond that, in most cases, only Python reference designs are available (marked with~*). It is worth noting that the Python reference design LoC is not fully indicative of the benchmark’s underlying complexity.

\subsection{Level 0}
Level-0 benchmarks are straightforward enough that all four LLMs consistently generate syntactically and functionally correct Verilog designs. However, for the \texttt{bitmanip\_unit} benchmark, which requires implementing one of four operations (rotate, mask, pack, and unpack), GPT-4.1 and DeepSeek R1 achieve only 70\% and 84\% correctness on the test cases, respectively, even in their best-performing generated codes. Appendix B, Table~\ref{tab:l0} presents the complete list of Level-0 benchmarks along with the number of lines of code in the reference designs. We specifically highlight three rows in the table where some LLMs fail to achieve a perfect score of 5 in the pass@5 metric.

\subsection{Level 1}
Among the Level-1a benchmarks (Table~\ref{tab:sixtybenchmarks}), the AES S-box implementation proved to be the most challenging task for the LLMs.. The S-box transforms an input byte into an output byte by applying a nonlinear substitution derived from finite field inversion followed by an affine transformation. There are two common implementation strategies for the S-box: one computes the substitution dynamically using logic, and the other precomputes a 256-entry lookup table (LUT) that directly maps inputs to outputs. In our evaluation, we specifically prompted the LLMs to generate a LUT-based implementation. Only Sonnet-4 and o4-mini-high were able to achieve 3/5 and 1/5 functionally correct codes, respectively. For the barrel shifter benchmark, DeepSeek R1 reached only 74\% correctness on the test cases in its best-performing outputs.

Level-1b (see Table~\ref{tab:sixtybenchmarks}) contains hierarchical adders, such as a 32-bit ripple carry adder built using 4-bit carry lookahead (CLA) blocks. More generally, these can be described as \((n, k)\)-bit adders, where \(n\) is the operand length and \(k\) is the width of each sub-adder block, opening up opportunities for design space exploration. All the evaluated LLMs were able to generate functionally correct code for these adders, as well as for the approximate adder benchmark. However, for the parameterized shift-and-add multiplier (with the unroll factor serving as the configurable parameter), the pass@5 results were 3/5 for Sonnet 4.0, 1/5 for GPT-4.1, 4/5 for o4-mini-high, and 1/5 for DeepSeek R1.

Level-1c contains complex arithmetic circuits. All the LLMs produced at least one functionally correct implementation for the Kogge-Stone and Brent-Kung adders, except DeepSeek R1, which achieved only 74.5\% test case correctness in its best-performing output. For the Booth multiplier, only o4-mini-high was able to generate Verilog code that passed all test cases. For the Wallace Tree multiplier, only DeepSeek R1 succeeded in generating a fully functional implementation. For the Dadda multiplier, all the LLMs failed to produce a correct design. In the restoring integer division benchmark, only GPT-4.1 and o4-mini-high generated codes that passed the complete testbench. Across all these cases, we also verified whether the generated code adhered to the specified architectural requirements. Even if a design was functionally correct, we classified it as incorrect if it followed an alternate architecture (for example, generating a regular multiplier instead of the specified Dadda multiplier).

\subsection{Level 2}
Level-2 contains pipelined versions of the arithmetic circuits introduced in Level-1c. Among the LLMs, o4-mini-high is able to generate correct designs for the pipelined versions of the ripple-carry adder (RCA), carry lookahead adder (CLA), and Wallace Tree multiplier, while DeepSeek R1 successfully produces a design for the pipelined CLA adder. In contrast to the lower levels, where syntax errors are relatively rare, we begin to observe a notable increase in syntax issues at higher levels. This is particularly evident in the outputs from Sonnet 4.0 and GPT-4.1 for the pipelined Wallace Tree multiplier, where many generated codes fail at compile time due to the complexity of the generated designs, which often rely on non-standard or invalid Verilog constructs. We note that our testbenches evaluate functional correctness only; performance metrics such as pipeline throughput are left for future work.

%We note that our testbenches evaluate functional correctness only, leaving performance metrics such as pipeline throughput to future work.

%It is important to note that our testbenches are designed to evaluate functional correctness only and do not measure pipeline throughput or performance — we leave such performance-focused evaluation to future work.

\subsection{Level 3}
Level-3 contains floating-point adder and multiplier circuits. Among the LLMs, Sonnet achieved 72\% and GPT~4.1 achieved 70\% on the floating-point adder benchmark. For the floating-point multiplier, Sonnet achieved 30\%, while GPT~4.1 achieved 60\%. The next group of benchmarks at this level focuses on iterative algorithms: Gauss-Seidel, Gradient Descent, the Newton-Raphson method for finding the root of a polynomial, and computing the square root of a real number. Notably, o4-mini-high successfully generates correct reference code for all these problems, while the other LLMs also show competitive performance — except for DeepSeek R1, which performs well only on the square root benchmark.

We believe this relatively strong performance on iterative algorithms is due to the precise mathematical formulation provided in their benchmark specifications, in contrast to circuits such as the Dadda multiplier, where the LLMs must rely on prior knowledge or fill in specification gaps from memory. For example, the design notes for the Gradient Descent benchmark explicitly provide the following equations:
\[
   f(x) = a x^2 + b x + c, \quad
   f'(x) = 2 a x + b, \quad
   x_{\text{next}} = x - \alpha \, f'(x)
\]
All arithmetic operations for these iterative algorithms are implemented using fixed-point representations.

\subsection{Levels 4, 5 and 6}
Levels 4, 5, and 6 feature highly complex, deeply pipelined datapaths, some incorporating floating-point arithmetic, for which all evaluated LLMs fail to produce functionally correct code when working solely from the provided problem descriptions and design specifications. Moreover, the generated code at these advanced levels exhibits syntax errors far more frequently than the smaller, simpler designs seen at the lower levels. Interestingly, Sonnet~4.0 is able to generate pipelined floating-point adder and multiplier designs that pass the complete testbench. GPT~4.1 and o4-mini-high also produce designs that pass over 90\% of the test cases. These results are somewhat surprising, given the models’ lower success rates on the iterative versions of these problems. This outcome may be attributed to clearer problem descriptions or the possibility that LLMs have been exposed to more pipelined designs during training. The other notable exception is the AES encryption core, where Sonnet 4.0 successfully generated a design that passed all our testbenches. We believe this success is likely due to AES being a widely known cryptographic accelerator, allowing the LLM to rely more heavily on its pretrained knowledge. Systolic arrays for matrix multiplication present another case where the code generated by the LLMs is able to pass the test cases, but primarily in simplified scenarios -- for example, when one of the matrices in the product is an identity matrix.

\begin{table}[tbp]
\centering
\scriptsize
\begin{adjustbox}{max width=\columnwidth}
\begin{tabular}{l l c >{\centering\arraybackslash}p{0.11\columnwidth} >{\centering\arraybackslash}p{0.11\columnwidth} >{\centering\arraybackslash}p{0.11\columnwidth} >{\centering\arraybackslash}p{0.11\columnwidth}}
\toprule
S.No. & Benchmark        & LoC & SON  & GPT  & O4M  & DSR  \\ \midrule

% Level-1a

\rowcolor{gray!20} \multicolumn{7}{l}{\textbf{Level-1a}} \\
1     & rca\_32bit        & 26   & \cellcolor{Green!80}(5, 100)  & \cellcolor{Green!80}(5, 100)  & \cellcolor{Green!80}(1, 100)   & \cellcolor{Green!80}(5, 100) \\
2     & cla\_8bit         & 56   & \cellcolor{Green!80}(5, 100)   & \cellcolor{Green!80}(5, 100)       & \cellcolor{Green!80}(5, 100)  & \cellcolor{Green!80}(5, 100) \\
3     & barrel\_shifter   & 128   & \cellcolor{Green!80}(5, 100)  & \cellcolor{Green!80}(4, 100)        & \cellcolor{Green!80}(4, 100)  & \cellcolor{YellowGreen!80}(1, 76) \\
4     & lfsr\_cipher      & 36   & \cellcolor{Green!80}(5, 100)  & \cellcolor{Green!80}(4, 100)       & \cellcolor{Green!80}(5, 100)   & \cellcolor{Green!80}(5, 100) \\
5     & gf\_field\_mult   & 60   & \cellcolor{Green!80}(3, 100)   & \cellcolor{Green!80}(5, 100)  & \cellcolor{Green!80}(3, 100)  & \cellcolor{Green!80}(2, 100) \\
6     & sbox\_lut         & 90   & \cellcolor{Green!80}(3, 100)  & \cellcolor{Red!60}(F, F)  & \cellcolor{Green!80}(1, 100)  & \cellcolor{Red!60}(F, F) \\

% Level-1b

\rowcolor{gray!20} \multicolumn{7}{l}{\textbf{Level-1b}} \\
7     & rca\_32bit\_cla          
& 70   
& \cellcolor{Green!80}(5, 100)
& \cellcolor{Green!80}(2, 100)
& \cellcolor{Green!80}(5, 100)  
& \cellcolor{Green!80}(5, 100) \\

8     & carry\_skip\_32bit       
& 55   
& \cellcolor{Green!80}(5, 100)  
& \cellcolor{Green!80}(5, 100)      
& \cellcolor{Green!80}(4, 100)  
& \cellcolor{Green!80}(3, 100) \\

9     & carry\_select\_32bit     
& 80   
& \cellcolor{Green!80}(5, 100)  
& \cellcolor{Green!80}(5, 100)      
& \cellcolor{Green!80}(5, 100)  
& \cellcolor{Green!80}(5, 100) \\

10     & shift\_add\_mult\_param  
& 126   
& \cellcolor{Green!80}(3, 100)  
& \cellcolor{Green!80}(1, 100)      
& \cellcolor{Green!80}(4, 100)  
& \cellcolor{Green!80}(1, 100) \\

11     & approx\_lower\_or\_adder 
& 31   
& \cellcolor{Green!80}(4, 100)  
& \cellcolor{Green!80}(2, 100)      
& \cellcolor{Green!80}(5, 100)  
& \cellcolor{Green!80}(4, 100) \\

% Level-1c

\rowcolor{gray!20} \multicolumn{7}{l}{\textbf{Level-1c}} \\

12     & kogge\_stone\_32bit      
& 106   
& \cellcolor{Green!80}(5, 100)  
& \cellcolor{Green!80}(4, 100)      
& \cellcolor{Green!80}(5, 100)  
& \cellcolor{Green!80}(3, 100) \\

13     & brent\_kung\_32bit       
& 158   
& \cellcolor{Green!80}(3, 100)  
& \cellcolor{Green!80}(4, 100)      
& \cellcolor{Green!80}(4, 100)  
& \cellcolor{YellowGreen!80}(2, 75) \\ %change color

14     & booth\_mult              
& 101   
& \cellcolor{orange!80}(5, 15)  
& \cellcolor{orange!80}(4, 15)      
& \cellcolor{Green!80}(5, 100)  
& \cellcolor{orange!80}(3, 11) \\ %change color

15    & wallace\_tree\_mult      
& 162   
& \cellcolor{Green!80}(4, 38)  
& \cellcolor{Green!80}(3, 34)      
& \cellcolor{Green!80}(4, 5)  
& \cellcolor{Green!80}(2, 100) \\ %change col

16    & dadda\_mult              
& 154  
& \cellcolor{orange!80}(2, 15)  
& \cellcolor{Red!60}(F, F)      
& \cellcolor{orange!80}(3, 11)  
& \cellcolor{orange!80}(4, 15) \\

17    & restoring\_div           
& 91   
& \cellcolor{Green!80}(4, 25)  
& \cellcolor{Green!80}(4, 100)      
& \cellcolor{Green!80}(3, 100)  
& \cellcolor{Red!60}(F, F) \\

18    & constant\_div            
& 22   
& \cellcolor{Green!80}(5, 100)  
& \cellcolor{Green!80}(5, 100)      
& \cellcolor{Green!80}(5, 100)  
& \cellcolor{Green!80}(5, 100) \\

% Level-2

\rowcolor{gray!20} \multicolumn{7}{l}{\textbf{Level-2}} \\

19    & aes128\_single\_encrypt  
& 192   
& \cellcolor{Green!80}(4, 100)  
& \cellcolor{Red!60}(4, F)      
& \cellcolor{Green!80}(4, 100)  
& \cellcolor{Red!60}(3, F) \\

20    & rca\_32bit\_pipe         
& 63   
& \cellcolor{yellow!80}(5, 30)  
& \cellcolor{orange!80}(5, 6)      
& \cellcolor{Green!80}(5, 100)  
& \cellcolor{orange!80}(4, 16) \\

21     & cla\_32bit\_pipe         
& 251 
& \cellcolor{orange!80}(5, 2)  
& \cellcolor{Red!60}(1, F)      
& \cellcolor{Green!80}(5, 100)  
& \cellcolor{Green!80}(4, 100) \\

22     & wallace\_tree\_mult\_pipe
& 208   
& \cellcolor{orange!80}(1, 24)  
& \cellcolor{orange!80}(1, 28)      
& \cellcolor{Green!80}(3, 100)  
& \cellcolor{Red!60}(1, F) \\

23    & dadda\_mult\_pipe        
& 224   
& \cellcolor{yellow!80}(3, 45)  
& \cellcolor{Red!60}(F, F)      
& \cellcolor{yellow!80}(3, 30)  
& \cellcolor{Red!60}(3, F) \\

% Level-3

\rowcolor{gray!20} \multicolumn{7}{l}{\textbf{Level-3}} \\
24    & fp\_adder       
& 88   
& \cellcolor{YellowGreen!80}(2, 72)  
& \cellcolor{YellowGreen!80}(1, 70)  
& \cellcolor{Red!60}(F, F)  
& \cellcolor{Red!60}(F, F) \\

25    & fp\_mult        
& 120   
& \cellcolor{orange!80}(2, 30)  
& \cellcolor{YellowGreen!80}(5,60)  
& \cellcolor{yellow!80}(2, 40)  
& \cellcolor{orange!80}(2, 30) \\

26    & gauss\_siedel   
& 102   
& \cellcolor{Green!80}(5, 100)  
& \cellcolor{YellowGreen!80}(5, 86)  
& \cellcolor{YellowGreen!80}(5, 90)  
& \cellcolor{Red!60}(1, F) \\

27    & grad\_desc      
& 79   
& \cellcolor{Green!80}(5, 100)  
& \cellcolor{Green!80}(4, 100)  
& \cellcolor{Green!80}(5, 100)   
& \cellcolor{Red!60}(F, F) \\

28    & nr\_poly        
& 165   
& \cellcolor{Red!60}(5, F)    
& \cellcolor{Red!60}(5, F) 
& \cellcolor{Green!80}(5, 100)  
& \cellcolor{Red!60}(5, F) \\

29    & nr\_sqrt        
& 76   
& \cellcolor{Green!80}(5, 100)  
& \cellcolor{YellowGreen!80}(5, 99.9)  
& \cellcolor{Green!80}(5, 100)  
& \cellcolor{Green!80}(4, 100) \\

% Level-4
\rowcolor{gray!20} \multicolumn{7}{l}{\textbf{Level-4}} \\

30  & fft\_16pt\_iter        
& 73*   
& \cellcolor{Red!60}(2, F)  
& \cellcolor{Red!60}(2, F)  
& \cellcolor{Red!60}(4, F)  
& \cellcolor{Red!60}(F, F) \\

31  & ifft\_16pt\_iter       
& 73*   
& \cellcolor{Red!60}(3, F)  
& \cellcolor{Red!60}(3, F)  
& \cellcolor{Red!60}(4, F)  
& \cellcolor{Red!60}(2, F) \\

32  & bandpass\_fir          
& 81*   
& \cellcolor{Red!60}(4, F)  
& \cellcolor{Red!60}(0, F)  
& \cellcolor{Red!60}(1, F)  
& \cellcolor{Red!60}(0, F) \\

33  & highpass\_fir          
& 72*   
& \cellcolor{Red!60}(4, F)  
& \cellcolor{Red!60}(0, F)  
& \cellcolor{Red!60}(1, F)  
& \cellcolor{Red!60}(F, F) \\

34  & lowpass\_fir           
& 67*   
& \cellcolor{Red!60}(5, F)  
& \cellcolor{Red!60}(1, F)  
& \cellcolor{Red!60}(2, F)  
& \cellcolor{Red!60}(F, F) \\

35  & fp\_adder\_pipe        
& 224   
& \cellcolor{Green!80}(4, 100)  
& \cellcolor{yellow!80}(1, 48)  
& \cellcolor{yellow!80}(4, 31)  
& \cellcolor{Orange!80}(1, 9) \\

36  & fp\_mult\_pipe         
& 161   
& \cellcolor{Green!80}(2, 100)  
& \cellcolor{Green!80}(3, 90)  
& \cellcolor{Green!80}(1, 97)  
& \cellcolor{Orange!80}(1, 27) \\

% Level-5
\rowcolor{gray!20} \multicolumn{7}{l}{\textbf{Level-5}} \\

37  & conv1d                 
& 78   
& \cellcolor{Red!60}(1, F)  
& \cellcolor{Red!60}(1, F)  
& \cellcolor{Red!60}(2, F)  
& \cellcolor{Red!60}(0, F) \\

38  & conv2d                 
& 128   
& \cellcolor{Red!60}(3 ,F)  
& \cellcolor{Red!60}(2, F)  
& \cellcolor{Red!60}(1, F)  
& \cellcolor{Red!60}(1, F) \\

39  & unsharp\_mask          
& 209   
& \cellcolor{Red!60}(4, F)  
& \cellcolor{Red!60}(3, F)  
& \cellcolor{Red!60}(4, F)  
& \cellcolor{Red!60}(1, F) \\

40  & harris\_corner         
& 173   
& \cellcolor{Red!60}(5, F)  
& \cellcolor{Red!60}(3, F)  
& \cellcolor{Red!60}(4, F)  
& \cellcolor{Red!60}(F, F) \\

41  & dct\_idct\_8pt\_1d     
& 68*   
& \cellcolor{Red!60}(4, F)  
& \cellcolor{Red!60}(1, F)  
& \cellcolor{Red!60}(3, F)  
& \cellcolor{Red!60}(1, F) \\

42  & systolic\_gemm         
& 90   
& \cellcolor{yellow!80}(4, 50\%)    
& \cellcolor{yellow!80}(5, 50\%)    
& \cellcolor{yellow!80}(5, 50\%)   
& \cellcolor{yellow!80}(5, 50\%)   \\

% Level-6
\rowcolor{gray!20} \multicolumn{7}{l}{\textbf{Level-6}} \\

43  & aes128\_encryption     
& 273  
& \cellcolor{Green!80}(5, 100)  
& \cellcolor{Red!60}(3, F)  
& \cellcolor{Red!60}(3, F)  
& \cellcolor{Red!60}(F, F) \\

44  & aes128\_decryption     
& 273  
& \cellcolor{Red!60}(3, F)  
& \cellcolor{Red!60}(4, F)  
& \cellcolor{Red!60}(3, F)  
& \cellcolor{Red!60}(F, F) \\

45  & fp\_bandpass\_fir      
& 81*  
& \cellcolor{Red!60}(5, F)  
& \cellcolor{Red!60}(F, F)  
& \cellcolor{Red!60}(F, F)  
& \cellcolor{Red!60}(F, F) \\

46  & fp\_highpass\_fir      
& 72*  
& \cellcolor{Red!60}(1, F)  
& \cellcolor{Red!60}(F, F)  
& \cellcolor{Red!60}(F, F)  
& \cellcolor{Red!60}(F, F) \\

47  & fp\_lowpass\_fir       
& 67*  
& \cellcolor{Red!60}(5, F)  
& \cellcolor{Red!60}(F, F)  
& \cellcolor{Red!60}(F, F)  
& \cellcolor{Red!60}(F, F) \\

48  & fft\_64pt\_pipelined   
& 71*  
& \cellcolor{Red!60}(1, F)  
& \cellcolor{Red!60}(3, F)  
& \cellcolor{Red!60}(2, F)  
& \cellcolor{Red!60}(F, F) \\

49  & conv3d                 
& 27*  
& \cellcolor{Red!60}(3, F)  
& \cellcolor{Red!60}(3, F)  
& \cellcolor{Red!60}(3, F)  
& \cellcolor{Red!60}(F, F) \\

50  & matmul\_quant         
& 34*  
& \cellcolor{Red!60}(F, F)  
& \cellcolor{Red!60}(3, F)  
& \cellcolor{Red!60}(4, F)  
& \cellcolor{Red!60}(5, F) \\

51  & conv2d\_multi\_channel 
& 25*  
& \cellcolor{Red!60}(4, F)  
& \cellcolor{Red!60}(2, F)  
& \cellcolor{Red!60}(4, F)  
& \cellcolor{Red!60}(1, F) \\

\bottomrule
\end{tabular}
\end{adjustbox}
\vspace{0.25em}
\caption{Performance of LLMs across 51 benchmarks using the pass@5 metric, including Lines of Code (LoC) in reference designs. An entry \((n, t)\) indicates \(n/5\) generations were syntactically correct, and among them, the best case passed \(t\%\) of the testbench. `F' indicates complete failure. Benchmarks using a Python reference design have LoC marked with an asterisk (*).
 }
\label{tab:sixtybenchmarks}
\end{table}

\section{Discussion}
\label{sec:discussion}
The evaluation across six benchmark levels reveals a clear complexity cliff for large language models (LLMs) in hardware design. While simple combinational circuits are often within reach, tasks involving deep pipelines, floating-point arithmetic, or bespoke architectures consistently defeat current models. 

Interestingly, the study finds that training-set familiarity plays a role: designs like AES, which are commonly encountered online, show occasional LLM success, whereas less common designs like Dadda multipliers see failures across models. Moreover, rich mathematical scaffolding in prompts (as with gradient descent) can significantly boost performance, underscoring that prompt quality, not just model size, influences outcomes. This observation raises a research question: could providing a clear algorithmic specification of the Dadda multiplier enable LLMs to leverage both their reasoning capabilities and training recall to generate functionally correct code? 

%This raises a research question: could providing a clear algorithmic specification of the Dadda multiplier help LLMs leverage both their reasoning and training recall to generate correct code?

%could providing a clear algorithmic specification of the Dadda multiplier enable LLMs to leverage both their reasoning capabilities and training recall to generate functionally correct code? 

To gain further insight into the above question, we explored the systolic array matrix multiplication benchmark. Along with the problem description and design specs, we asked GPT~4.1 to first generate C code easily translatable to HDL. A follow-up prompt then produced Verilog code that computes the matrix product correctly—except for the last row. We believe this issue could be fixed with reflective refinement. This small experiment hints at agentic approaches we might adopt for complex digital system design.

\section{Conclusions}
\label{sec:conclusions}
\needspace{2\baselineskip}
\begin{samepage}
\emph{ArchXBench} addresses an existing gap by providing benchmarks with appropriate architectural complexity that reflect 
practical designs, along with their associated area, latency and throughput requirements. The GitHub repository, which we plan to make public, includes clear problem descriptions, design specifications, and validated testbenches.We believe this will help advance research on LLM-driven agentic approaches for complex digital system design. We invite the community to build on this suite and extend it with new benchmarks, advanced design tasks, and \nopagebreak\ 
 richer evaluation protocols to help guide next generation automated SoC design tools.
\end{samepage}
\nopagebreak

\iffalse

\emph{ArchXBench} closes a long-standing gap in HDL evaluation by coupling natural-language task descriptions with formal interfaces, testbenches, and reference models that scale from trivial adders to full AES and CNN subsystems.
A zero-shot study of four commercial LLMs shows near-perfect success on Level 0 yet frequent architectural drift, syntax errors, and outright failures once pipelining, floating-point, or hierarchical reuse appear—evidence of a steep complexity cliff that today’s code-first models cannot bridge alone.
Because every benchmark is self-contained and open-sourced, ArchXBench now offers a reproducible yardstick for measuring retrieval-augmented, agentic, or ensemble approaches that aim to deliver production-ready RTL.
We invite the community to build on this suite, extend it with area- and timing-aware checks, and use its fine-grained failure modes to guide the next generation of automated SoC design tools.

\fi 
%\newpage
\begin{balance}
\bibliography{references}
\bibliographystyle{abbrv}
\end{balance}

\hspace{1cm}
\newline
\newpage 
\clearpage

\pagebreak
\newpage
\section{Appendix A}
\label{sec:appendix}
Example prompt, design specifications, Python reference design, and Verilog testbench from the Level 6 3D convolution accelerator benchmark. The benchmark also includes (not shown here) Python scripts for input generation, output evaluation, and Makefiles for workflow orchestration.

%%%%%%%%%%%%%%%%%%%%% Start of listing 1 %%%%%%%%%%%%%%%%%%%%%%%%%%%%%%
\vspace{0.5em}

\noindent\textbf{Listing 1.} Prompt for the 3D convolution accelerator benchmark.

\begin{tcolorbox}[
  colback=white!95!gray,
  colframe=black!75!black,
  width=\dimexpr\columnwidth\relax, % ensures it matches one column
  arc=2mm,
  boxrule=0.5pt,
  fontupper=\footnotesize,
  breakable,
  enhanced jigsaw
]
\textbf{Prompt: 3D Convolution — Streaming Accelerator for Spatiotemporal Filtering}

\textbf{Objective:} Design a pipelined hardware accelerator for performing 3D convolution on a stream of 3D image or video data. The convolution must apply a 3D kernel over a 3D input volume (e.g., height × width × time or depth), producing a filtered output volume. This benchmark stresses spatial and temporal data buffering, window generation, and MAC compute reuse.

\textbf{Background:} 3D convolutions are essential in computer vision, video understanding, volumetric medical imaging (CT/MRI), and spatiotemporal deep learning. A 3D convolution involves scanning a 3D kernel across a 3D input volume and computing a sum of elementwise products at each valid position. Unlike 2D convolution, 3D convolution has one more axis (e.g., time or depth), significantly increasing the data reuse complexity.

\textbf{Design Constraints:}
\begin{itemize}
    \item Inputs: a stream of 3D data values (e.g., voxel grid or video frames).
    \item Kernel size: typically 3×3×3 or 5×5×5, parameterizable.
    \item Padding and stride handling may be included or assumed fixed.
    \item All computations must be pipelined; 1 output voxel per cycle after filling.
    \item Intermediate buffering must support 3D stencil window formation. 
\end{itemize}

\textbf{Performance Expectation:} After pipeline fill, the design must produce one valid output per cycle. Efficient reuse of 3D data slices (via shift registers or FIFO arrays) is key to minimizing bandwidth and maximizing throughput.

\textbf{Deliverables:}
\begin{itemize}
    \item Parameterized Verilog/HLS implementation with streaming I/O.
    \item Separate line/frame/volume buffers for 3D sliding window generation.
 %   \item Testbench with synthetic 3D volumes and golden reference outputs.

\end{itemize}
\end{tcolorbox}

%\vspace{0.5em}
%\noindent\textbf{Listing 1.} Prompt for the 3D convolution accelerator benchmark.

%%%%%%%%%%%%%%%%%%%%% End of listing 1 %%%%%%%%%%%%%%%%%%%%%%%%%%%%%%

%%%%%%%%%%%%%%%%%%%%% Start of listing 2 %%%%%%%%%%%%%%%%%%%%%%%%%%%%%%

\vspace{3em}
\noindent\textbf{Listing 2.} Level 6 design specifications for the 3D convolution accelerator benchmark.

\begin{tcolorbox}[
  colback=white!95!gray,
  colframe=black!75!black,
  width=\dimexpr\columnwidth\relax,
  arc=2mm,
  boxrule=0.5pt,
  fontupper=\footnotesize\ttfamily,
  breakable,
  enhanced jigsaw,
]
Design Name: conv3d\_streaming\_accelerator \\
Module Name: conv3d \\

Inputs: \\
- clk\hspace*{25mm}// Clock signal \\
- rst\hspace*{25mm}// Active-high reset \\
- voxel\_in[DATA\_W-1:0]\hspace*{4mm}// Input 3D voxel stream (flattened) \\
- valid\_in\hspace*{20mm}// Input valid signal \\
- kernel[K3*K2*K1*DATA\_W-1:0]\hspace*{2mm}// Flattened 3D kernel coefficients \\
- last\_in\hspace*{21mm}// End-of-volume marker \\

Outputs: \\
- voxel\_out[DATA\_W+LOG\_KW-1:0]\hspace*{2mm}// Output voxel after convolution \\
- valid\_out\hspace*{20mm}// Output valid signal \\
- done\hspace*{27mm}// End-of-volume signal \\

Parameters: \\
- K1, K2, K3\hspace*{15mm}// Kernel depth × height × width \\
- D, H, W\hspace*{18mm}// Input volume dimensions (Depth, Height, Width) \\
- DATA\_W\hspace*{23mm}// Bit-width of input/output voxels \\

Design Signature: \\

module conv3d \#( \\
\hspace*{2em}parameter K1 = 3,  // Kernel depth \\
\hspace*{2em}parameter K2 = 3,  // Kernel height \\
\hspace*{2em}parameter K3 = 3,  // Kernel width \\
\hspace*{2em}parameter D = 8,   // Input volume depth \\
\hspace*{2em}parameter H = 64,  // Input height \\
\hspace*{2em}parameter W = 64,  // Input width \\
\hspace*{2em}parameter DATA\_W = 8 \\
) ( \\
\hspace*{2em}input clk, \\
\hspace*{2em}input rst, \\
\hspace*{2em}input [DATA\_W-1:0] voxel\_in, \\
\hspace*{2em}input valid\_in, \\
\hspace*{2em}input [K1*K2*K3*DATA\_W-1:0] kernel, \\
\hspace*{2em}input last\_in, \\
\hspace*{2em}output [DATA\_W+LOG\_KW-1:0] voxel\_out, \\
\hspace*{2em}output valid\_out, \\
\hspace*{2em}output done \\
); \\

Design Notes: \\
- The input data must be buffered in 3D using: \\
\hspace*{2em}• Line buffers (for rows), \\
\hspace*{2em}• Frame buffers (for slices), \\
\hspace*{2em}• Depth buffers (for temporal storage). \\

- Window formation: \\
\hspace*{2em}• Form a 3D sliding window of size K1×K2×K3 from the buffered data. \\
\hspace*{2em}• Multiply each voxel in the window with the corresponding kernel weight. \\

- MAC Array: \\
\hspace*{2em}• Use a single pipelined 3D MAC tree or parallel units per voxel. \\
\hspace*{2em}• Intermediate results may be accumulated using wide registers. \\

- Stride = 1; padding = valid; optionally parameterized. \\
- Final output may be scaled, clamped, or truncated to fit output width. \\

Optional Extensions: \\
- Stride/padding control \\
- Pipelined multi-channel 3D conv (e.g., input/output feature maps)
\end{tcolorbox}

%%%%%%%%%%%%%%%%%%%%% End of listing 2 %%%%%%%%%%%%%%%%%%%%%%%%%%%%%%

%\newline  

\newpage
%%%%%%%%%%%%%%%%%%%%% Start of listing 3 %%%%%%%%%%%%%%%%%%%%%%%%%%%%%%

\vspace{0.5em}
\noindent\textbf{Listing 3.} Level 6 Python reference design for the 3D convolution accelerator benchmark.

% Define Python style for listings
\lstdefinestyle{pythonstyle}{
  language=Python,
  basicstyle=\ttfamily\footnotesize,
  keywordstyle=\color{blue},
  commentstyle=\color{gray},
  stringstyle=\color{red!70!black},
  showstringspaces=false,
  breaklines=true,
  frame=none,
}

% In the document body (inside onecolumn)
\begin{tcolorbox}[
  colback=white!95!gray,
  colframe=black!75!black,
  width=\dimexpr\columnwidth\relax,
  arc=2mm,
  boxrule=0.5pt,
  breakable,
  enhanced jigsaw,
]
\begin{lstlisting}[style=pythonstyle]
import json
import numpy as np

def conv3d(volume, kernel):
    D, H, W = volume.shape
    kD, kH, kW = kernel.shape
    out_D = D - kD + 1
    out_H = H - kH + 1
    out_W = W - kW + 1
    output = np.zeros((out_D, out_H, out_W), dtype=np.int32)

    for d in range(out_D):
        for i in range(out_H):
            for j in range(out_W):
                patch = volume[d:d+kD, i:i+kH, j:j+kW]
                output[d, i, j] = np.sum(patch * kernel)
    return output

with open("inputs/stimuli.json") as f:
    data = json.load(f)

vol = np.array(data["volume"], dtype=np.int32)
kernel = np.ones((3,3,3), dtype=np.int32)
C = conv3d(vol, kernel)

with open("outputs/golden_output.json", "w") as f:
    json.dump({"C": C.flatten().tolist()}, f, indent=2)
\end{lstlisting}
\end{tcolorbox}

%%%%%%%%%%%%%%%%%%%%%%%%%%End of listing 3 %%%%%%%%%%%%%%%%%%%%%%%%%%%%%%%%%%%%%%%%%%%%%%

%%%%%%%%%%%%%%%%%%%%% Start of listing 4 %%%%%%%%%%%%%%%%%%%%%%%%%%%%%%
\vspace{3em}
\noindent\textbf{Listing 4.} Level 6 Verilog testbench for the 3D convolution accelerator benchmark.
\begin{tcolorbox}[
  colback=white!95!gray,
  colframe=black!75!black,
  width=\dimexpr\columnwidth\relax,
  arc=2mm,
  boxrule=0.5pt,
  breakable,
  enhanced jigsaw,
]
\begin{lstlisting}[style=verilogstyle]
`timescale 1ns/1ps

module tb_conv3d;

  // Parameters matching DUT
  parameter K1 = 3, K2 = 3, K3 = 3;
  parameter D  = 8, H  = 64, W  = 64;
  parameter DATA_W = 8;

  // Compute total number of input voxels
  localparam N     = D * H * W;
  localparam OUT_W = DATA_W + 4;  // DATA_W + LOG_KW 

  // Testbench signals
  reg                      clk, rst, valid_in, last_in;
  reg  [DATA_W-1:0]        voxel_in;
  reg  [K1*K2*K3*DATA_W-1:0] kernel;
  wire [OUT_W-1:0]         voxel_out;

  // Memory for loading input volume
  reg [7:0] input_volume [0:N-1];
  integer i, fout;

  // Instantiate DUT
  conv3d #(
    .K1(K1), .K2(K2), .K3(K3),
    .D(D),  .H(H),  .W(W),
    .DATA_W(DATA_W)
  ) dut (
    .clk(clk),
    .rst(rst),
    .voxel_in(voxel_in),
    .valid_in(valid_in),
    .kernel(kernel),
    .last_in(last_in),
    .voxel_out(voxel_out),
    .valid_out(),  // not used in dummy DUT
    .done()        // not used
  );

  // Generate clock (10 ns period)
  always #5 clk = ~clk;

  initial begin
    // Load input volume from memory file
    $readmemh("tb_input.mem", input_volume);

    // Initialize signals
    clk      = 0;
    rst      = 1;
    valid_in = 0;
    last_in  = 0;
    kernel   = 0;  // dummy kernel

    #20 rst = 0;

    // Ensure outputs directory exists (via Makefile's run target)

    // Open output JSON file
    fout = $fopen("outputs/dut_output.json", "w");
    $fwrite(fout, "{\n  \"C\": [\n");

    // Stream inputs and capture outputs
    for (i = 0; i < N; i = i + 1) begin
      @(negedge clk);
      voxel_in = input_volume[i];
      valid_in = 1;
      last_in  = (i == N-1);
      @(posedge clk);
      // write this cycle's output
      $fwrite(fout, "    %0d%s\n",
              voxel_out,
              (i == N-1) ? "" : ",");
    end

    // Close JSON array and file
    $fwrite(fout, "  ]\n}\n");
    $fclose(fout);

    $display("[TB] Simulation complete, outputs/dut_output.json written.");
    #10 $finish;
  end

endmodule
\end{lstlisting}
\end{tcolorbox}

%%%%%%%%%%%%%%%%%%%%% End of listing 4 %%%%%%%%%%%%%%%%%%%%%%%%%%%%%%
\newpage
\section{Appendix B}

%Note that reference designs are not included in the benchmark to avoid LLMs from incorporating this into their training. 

\begin{table}[htbp]\label{tab:l0}
  \centering
  \begin{tabular}{r l r c c c c}
    \toprule
    S.No. & Circuit             & LoC & SON                & GPT                & O4M                & DSR                \\
    \midrule
    1  & binary-encoder       & 12  & \cellcolor{Green!80}(2, 100)  & \cellcolor{Green!80}(4, 100)  & \cellcolor{Green!80}(5, 100)  & \cellcolor{Green!80}(4, 100)  \\
    2  & bitmanip-unit        & 60  & \cellcolor{Green!80}(5, 100)  & \cellcolor{YellowGreen!80}(5, 70)  & \cellcolor{Green!80}(5, 100)  & \cellcolor{YellowGreen!80}(5, 84)  \\
    3  & clock-divide-by-2    & 19  & \cellcolor{Green!80}(5, 100)  & \cellcolor{Green!80}(5, 100)  & \cellcolor{Green!80}(5, 100)  & \cellcolor{Green!80}(5, 100)  \\
    4  & comparator-4bit      & 5   & \cellcolor{Green!80}(5, 100)  & \cellcolor{Green!80}(5, 100)  & \cellcolor{Green!80}(5, 100)  & \cellcolor{Green!80}(5, 100)  \\
    5  & comparator-8bit      & 46  & \cellcolor{Green!80}(5, 100)  & \cellcolor{Green!80}(5, 100)  & \cellcolor{Green!80}(3, 100)  & \cellcolor{Green!80}(2, 100)  \\
    6  & decoder-2to4         & 19  & \cellcolor{Green!80}(4, 100)  & \cellcolor{Green!80}(3, 100)  & \cellcolor{Green!80}(2, 100)  & \cellcolor{Green!80}(5, 100)  \\
    7  & decoder-3to8         & 33  & \cellcolor{Green!80}(5, 100)  & \cellcolor{Green!80}(5, 100)  & \cellcolor{Green!80}(5, 100)  & \cellcolor{Green!80}(5, 100)  \\
    8  & demux-1to2           & 14  & \cellcolor{Green!80}(5, 100)  & \cellcolor{Green!80}(5, 100)  & \cellcolor{Green!80}(5, 100)  & \cellcolor{Green!80}(5, 100)  \\
    9  & demux-1to4           & 30  & \cellcolor{Green!80}(5, 100)  & \cellcolor{Green!80}(5, 100)  & \cellcolor{Green!80}(5, 100)  & \cellcolor{Green!80}(5, 100)  \\
    10 & down-counter         & 23  & \cellcolor{Green!80}(5, 100)  & \cellcolor{Green!80}(5, 100)  & \cellcolor{Green!80}(5, 100)  & \cellcolor{Green!80}(5, 100)  \\
    11 & gray-counter         & 19  & \cellcolor{Green!80}(5, 100)  & \cellcolor{Green!80}(5, 100)  & \cellcolor{Green!80}(5, 100)  & \cellcolor{Green!80}(5, 100)  \\
    12 & johnson-counter      & 15  & \cellcolor{Green!80}(5, 100)  & \cellcolor{Green!80}(5, 100)  & \cellcolor{Green!80}(5, 100)  & \cellcolor{Green!80}(5, 100)  \\
    13 & mux2to1              & 17  & \cellcolor{Green!80}(5, 100)  & \cellcolor{Green!80}(5, 100)  & \cellcolor{Green!80}(5, 100)  & \cellcolor{Green!80}(5, 100)  \\
    14 & mux4to1              & 40  & \cellcolor{Green!80}(5, 100)  & \cellcolor{Green!80}(5, 100)  & \cellcolor{Green!80}(5, 100)  & \cellcolor{Green!80}(5, 100)  \\
    15 & priority-encoder     & 14  & \cellcolor{Green!80}(5, 100)  & \cellcolor{Green!80}(5, 100)  & \cellcolor{Green!80}(5, 100)  & \cellcolor{Green!80}(5, 100)  \\
    16 & ring-counter         & 12  & \cellcolor{Green!80}(5, 100)  & \cellcolor{Green!80}(5, 100)  & \cellcolor{Green!80}(5, 100)  & \cellcolor{Green!80}(5, 100)  \\
    17 & sipo-regs            & 20  & \cellcolor{Green!80}(5, 100)  & \cellcolor{Green!80}(5, 100)  & \cellcolor{Green!80}(5, 100)  & \cellcolor{Green!80}(5, 100)  \\
    18 & siso-regs            & 20  & \cellcolor{Green!80}(5, 100)  & \cellcolor{Green!80}(5, 100)  & \cellcolor{Green!80}(5, 100)  & \cellcolor{Green!80}(4, 100)  \\
    19 & up-counter           & 18  & \cellcolor{Green!80}(5, 100)  & \cellcolor{Green!80}(5, 100)  & \cellcolor{Green!80}(5, 100)  & \cellcolor{Green!80}(5, 100)  \\
    20 & piso-regs           & 18  & \cellcolor{Green!80}(5, 100)  & \cellcolor{Green!80}(5, 100)  & \cellcolor{Green!80}(5, 100)  & \cellcolor{Green!80}(5, 100)  \\
    \bottomrule
  \end{tabular}
  \vspace{0.5em}

  \caption{Complete list of Level 0 benchmarks and associated lines of code in the reference Verilog design. }
\end{table}

%%%%%%%%%%%%%%%%%%%%% End of listing 2 %%%%%%%%%%%%%%%%%%%%%%%%%%%%%%
\end{document}